\documentstyle[fleqn]{article}

\begin{document}

\title{Finite Precision Measurement Nullifies Euclid's Postulates}

\author{Asher Peres}
\date{Department of Physics, Technion---Israel Institute of
Technology, 32000 Haifa, Israel}

\maketitle
\begin{abstract}
Following Meyer's argument [Phys. Rev. Lett. {\bf 83}, 3751 (1999)]
the set of all directions in space is replaced by the dense subset of
rational directions. The result conflicts with Euclidean geometry.
\end{abstract}

\bigskip

Meyer's claim \cite{meyer} that ``finite precision measurement nullifies
the Kochen-Specker theorem'' (that is, makes it irrelevant to physics)
and some of its generalizations \cite{kent} have caused considerable
controversy that lasts until today \cite{arxiv}. Meyer's proposal
was to replace the set of all directions in space by the dense subset
of rational directions, arguing that a finite precision measurement
cannot decide whether or not a number is rational.

Let us apply the same argument to ordinary geometry and consider only
points with rational coordinates. Then the line $x=y$ and the unit
circle $x^2+y^2=1$ are both dense but they do not intersect, in
contradiction to Euclid's postulates \cite{euclid}.

This work was supported by the Gerard Swope Fund.

\end{document}